\documentstyle[11pt,aaspp4]{article}
\begin{document}
\received{}
\accepted{}
\lefthead{ }
\righthead{}
\title{HOW SATURATED ARE ABSORPTION LINES IN THE BROAD ABSORPTION
       LINE QUASAR PG 1411+442 ?}
\author{T.G. WANG,\altaffilmark{1,2} J.X. WANG,\altaffilmark{2}, 
W. BRINKMANN,\altaffilmark{3} M. MATSUOKA \altaffilmark{1}}
\altaffiltext{1}{The Institute of Physical and Chemical Research ({\it RIKEN}),
2-1, Hirosawa, Wako, Saitama, 351-0198, Japan}
\altaffiltext{2}{Center for Astrophysics, University of Science and Technology 
of China, Anhui, 230026, China}
\altaffiltext{3}{Max-Planck-Institut f\"ur extraterrestrische Physik, 
Giessenbachstrasse,D-85740 Garching, FRG}
\authoremail{wang@crab.riken.go.jp}

\begin{abstract}
Recently, convincing evidence was found for extremely large X-ray absorption
by column densities $>$ 10$^{23}$~cm$^{-2}$ in broad absorption line quasars. 
One consequence of this is that any soft X-ray emission from these QSOs would 
be the scattered light or leaked light from partially covering absorbing 
material. A detection of the unabsorbed soft X-ray and absorbed hard X-ray component 
will allow to determine the total column density as well as the effective 
covering factor of the absorbing material, which can be hardly obtained 
from the UV absorption lines. Brinkmann et al. (1999) showed that both the 
unabsorbed and absorbed components are detected in the nearby very bright 
broad absorption line quasar PG 1411+442. In this letter, we make a further 
analysis of the broad band X-ray spectrum and the UV spectrum from HST,  and 
demonstrate that broad absorption lines are completely saturated at the 
bottom of absorption troughs. 
\end{abstract}


\section{INTRODUCTION}
 
About  10\% of quasars (QSO) found in the optical flux-limited samples
exhibit broad absorption lines (BAL), resonance line absorption troughs
extending $\sim$ 0.1c to the blue of the emission line centers (Weymann et
al. 1991). However, Goodrich (1997) and Krolik \& Voit (1998) argued that
the true fraction of BAL QSO can be as high as $>30$\% considering  
attenuation of the light or a non-spherical distribution of continuum 
emission. Based on the
similarity of emission line properties, it is generally thought that
BAL regions exist in all quasars, but occupy only a fraction of the solid 
angle (e.g., Weymann et al. 1991). Thus a BAL region is an 
important part of every QSO's structure.

The absorption line properties of BAL QSO have been intensively studied,
however,  results on  the nature of the absorbing material from
the UV absorption lines alone are rather controversial.  All earlier results
indicated low column densities of about an equivalent $N_H$ density around
10$^{20\sim 21}$~cm$^{-2}$ on the assumption that the absorption is not too optically 
thick, but metal abundances far higher than the solar values are required 
(Korista et al. 1996; Turnshek et al. 1996; Hamann 1996). However, recent results
show that the structure of the absorbing material 
 is rather complicated and seriously saturated in some velocity range
in PG 0946+301 (Arav et al. 1998). The total column density could be much
larger. One major difficulty, however, is the lack of an independent measurement
 of the covering factor.
 
In the soft X-ray band, BAL QSO are notorious for their weakness (Green et 
al. 1995, Green \& Mathur 1996).
It was argued that the BAL QSO are not intrinsically X-ray weak,
but heavily absorbed in the soft band. Great similarity
of the  emission line spectra in BAL and non-BAL QSO indicates that the BLR
clouds have seen a similar ionizing (UV to X-ray) continuum in both type of
QSO (Weymann et al. 1991).

Heavy absorption has been found in the one of two ROSAT detected BAL
QSO.  Green \& Mathur (1996) showed that the spectrum of 1246-057 is
absorbed by an intrinsic column of $\sim$1.2$\times$10$^{23}$~cm$^{-2}$. A
similar column density was found in the ASCA spectrum of PHL 5200 (Mathur
et al. 1996). More BAL QSO still remain to be discovered in the X-ray
band. From the ROSAT non-detection, Green \& Mathur (1996) derived a lower
limit for the absorption column densities of $\sim$ 10$^{23}$~cm$^{-2}$ for
these objects if the intrinsic X-ray emission is similar to other quasars.
A similar conclusion has been reached  by
Gallagher et al. (1999). 
Brinkmann et al. (1999) analyzed the ASCA spectra of three BAL QSO; only the
brightest, low redshift (z=0.0896, Marziani et al. 1996) BAL QSO PG 1411+442 has been detected with a column 
density of $\sim$2~10$^{23}$~cm$^{-2}$. 

At these column densities, photons below 1 keV are completely absorbed. 
Any soft X-ray emission from a BAL QSO must be either scattered light 
or the light leaking from a partially covering absorber. Therefore, the 
measurement of the unabsorbed component in soft X-rays and the absorbed 
ones in the hard X-ray band will enable us to determine the fraction of light 
scattered or leaked, thus providing an independent measurement of the effective 
covering factor of absorbing material. Both, scattered light and absorbed 
hard X-rays were detected in the nearby bright BAL QSO PG 1411+442 (Brinkmann 
et al. 1999).       
 
In this {\it Letter}, we make a further analysis of the broad band X-ray 
spectrum and the UV absorption line spectrum from  HST. We will 
show that the UV absorption lines are completely saturated in the deepest 
part of the absorption trough.  

\section{ X-RAY ABSORPTION AND SCATTERING}

Brinkmann et al. (1999) found that the combined ROSAT and ASCA spectrum 
can be well fitted by a heavily absorbed power law at high energies  
plus an unabsorbed power law at low energy with a much steeper spectrum. 
They noticed that the steepness of the power law in the soft band is consistent 
with the H$\beta$ width versus soft X-ray spectral index correlation 
found for a sample of QSOs (Laor et 
al. 1997, Wang et al. 1996), and interpreted this component as scattered 
or leaked nuclear light. By comparing the normalizations at 1 keV, they 
estimated the fraction of this soft component to the primary component 
to be $\sim$ 5 per cent. Here we re-fit the broad band X-ray spectrum 
to determine more consistently the fraction of the scattered or 
leaked light.

As the ASCA and ROSAT spectra in the overlapping energy range are consistent with 
each other, they  were fitted simultaneously (see Brinkmann et al. 1999). Since 
the X-ray spectra at low energies are usually much steeper due to  
a soft excess, the X-ray spectrum is modeled 
as a broken power law with a low energy index $\Gamma_{sx}$ and a high 
energy index $\Gamma_{hx}$, with  
partially covering absorption. The index $\Gamma_{hx}$ is fixed to 
2.0, which is typical for radio-quiet QSOs. The model can fit the data 
very well ($\chi^2=74$ for 79 degrees of freedom, see also Figure 1a). The results are presented 
in Table 1. Figure 2 shows the 68\% and 90\% confidence contours for 
$N_H^{(intrin)}$ versus the covering factor. Both covering factor and 
the absorption column density are dependent on the value of $\Gamma_{hx}$ 
and the break energy. For a reasonable range  
$1.8<\Gamma_{hx}<2.2$, the best fit covering factor is around 0.95 to 0.97.  
The fitted covering factor also depends on the break energy. A lower break 
energy yields a slightly larger covering factor and a higher break energy results in a somewhat 
lower covering factor. Considering these uncertainties, it is likely that
the covering factor is in the range 0.94-0.97.

 Since PG 1411+442 is the least luminous BAL QSO, there is some
 concern that the 
soft X-ray emission might be due to thermal emission from hot gas in the 
host galaxy. A model consisting of an absorbed power-law plus thermal emission 
(Raymond-Smith model) was fitted to the spectra. The fit is statistically poor 
( $\chi^2$=132 for 78 d.o.f.) and shows systematic deviations 
in the residuals (Figure 1b). Furthermore, the fit yields a very flat photon 
 index $\Gamma=-0.06_{-0.15}^{+0.20}$.  Therefore, thermal 
emission cannot be the source  for the soft excess.  

As warm absorption can also produce a structure similar to a soft excess, 
we fit the spectrum with a power-law absorbed by ionized absorption 
material ( Zdziarski et al. 1995). The fit is only acceptable at a probability 
of 3\% ($\chi^2$=104 for 78 d.o.f.) with systematic deviations between 
0.3 and 1.2 keV (see Figure 1c). Thus a warm absorber cannot successfully 
reproduce the observed spectrum either. In fact, when a partially covered 
warm absorption model applied, the soft X-ray emission almost entirely 
come from the leaked component and the X-ray ionization parameter ($U_x$, 
see Netzer 1996 for definition) is constrainted to be 0.05$_{-0.05}^{+0.04}$.   

\section{THE UV ABSORPTION LINE TROUGHS}

PG 1411+442 was observed by HST with GHRS centered at 130nm 
and 190nm with a spectral resolution $\Delta\lambda/\lambda=3000$ (Corbin 
\& Boroson 1996). 
We retrieved the calibrated files from the HST data archive 
to obtain the wavelength, absolute flux, error flag, and the photon noise 
vector. We have averaged all spectra. Since the blue wings of the emission lines  
Ly$\alpha$, CIV and NV are strongly absorbed, it is impossible to determine 
their emission line profile. CIII]$\lambda$1909 is the only moderately strong 
line which is not affected by absorption in the HST spectrum, and is used 
to construct a template model for emission line profiles.  
We fitted the line with two gaussians over a range avoiding the 
SiIII] contamination and assumed that the other lines also consist of the 
same two gaussian components but with different normalizations. To test this, 
we have fitted this model to the HeII and the NV, CIV red wings. Reasonable 
good fits were obtained and normalizations are determined for each line. We 
have taken into account the doublet nature of CIV and NV line.  
The double ratios were fixed to the ratio of their statistical weights. 
Since the red wing 
of $Ly\alpha$ is also affected by the NV absorption line, the emission line 
profiles are modeled using the scaled exact model for CIV line profile.
 
Figure 3 shows the ratio of the data to a  model consisting of emission lines 
and a continuum. 
The absorption line profiles are similar for CIV, Ly$\alpha$ and NV. The 
difference in the apparent broadness of NV, CIV, and Ly$\alpha$ of the  
component B is due to the doublet nature of CIV and NV, and their different 
separations.  The narrow (FWHM $\sim$ 300 km~sec$^{-1}$) absorption line 
at zero velocity which is seen strongest in Ly$\alpha$ and is also visible 
in CIV and SiIV as doublets. The exact profile of component A in NV is 
uncertain due to the fact that the Ly$\alpha$ profile cannot be well defined. 
A more detailed analysis 
of the absorption line structures is beyond the scope of this paper.

The absorption troughs of the component B in CIV and NV are box-shaped 
and have a none-zero flux of  $\sim$ 5-6 per cent of the unabsorbed flux 
at the bottom. This value is 
higher than the expected Grating-scattered light at 1-2\% level in the GHRS 
spectrum (Crenshaw et al. 1998). However, the fraction of light at the 
bottom of the troughs  is coincident with the 
3-5 per cent of scattered or leaked light in the soft X-ray band.  

\section {DISCUSSION}

We have shown that in PG 1411+442 unabsorbed X-rays occur at a 3-5 per 
cent level, which is coincident with the fraction of the residual flux
in the bottom of the absorption line troughs. This can be described by either 
a partial coverage or a simple scattering model. In either case, 
the absorption lines at the bottom of troughs can quite well be black 
saturated.    
  
The idea that scattered light fills in the absorption line troughs in BAL 
QSOs was actually proposed some years ago based on the results of 
spectropolarimetric observations. Cohen et al. (1995) and Ogle (1997) showed 
that the polarization at the bottom of the troughs is much higher than the 
continuum level polarization. Their finding 
demonstrates that a considerable  fraction of the 
flux at the bottom of the troughs is from 
scattered photons. As the polarization degree is strongly dependent on the 
detailed geometry of the scatterer as well as on the continuum emission 
pattern, it is impossible to determine the amount of total scattered light 
from the polarized light itself and the degree of saturation of the line cannot 
be estimated. 
Here we have illustrated that because of the completely scattered or leaked 
nature of the soft X-rays, broad band X-ray observations can provide a 
way to estimate this fraction. A combination of polarization and X-ray 
measurements will allow to constrain the geometry of the scatterer. 

Ogle et al. (1997) showed that,in general, the absorbing medium also partially 
covers the scattering medium. This implies that the scattered light 
is also partially absorbed.  We wish to point out that the upper
limit ( 3.3$\times 10^{20}$~cm$^{-2}$) of column density derived for the 
intrinsic soft X-ray absorption might still be consistent with their results.
Future UV spectro-polarimetric observations of PG 1411+442 will allow to address
this question.  The remarkable good 
agreement between the ROSAT and ASCA spectrum at the
overlapping region suggests that the soft X-ray emission did not change between
the two epochs separated by  
6 years. This can be naturally explained in a scattering model.  
  
On the other hand, partial covering is strongly suggested from optical and 
UV observations. 
For intrinsic absorber with relatively narrow UV line profiles, there is good 
evidence that the coverage of the continuum source is partial, velocity 
dependent, and may also be a function of ionization stage (Barlow et al. 1997) 
. There 
is suspicion that the same complications apply to the BAL lines (Arav 1997).
A constant soft X-ray flux  is possibly a drawback for the partial covering model 
as the continuum of PG 1411+442 is variable (Giveon et al. 1999). Because of 
lacking simultaneous hard X-ray observation during the ROSAT observation, 
we are unable to rule out the partial covering model. Future observations of this 
object covering both, soft and hard X-rays,
will allow to discriminate the two models and to precisely estimate the 
fraction 
of unabsorbed X-ray light.

The X-ray properties of objects with intrinsically relatively narrow absorption 
lines are completely different from those of BAL QSOs. Warm absorption is seen 
in the former, while X-rays are extremely weak in the BAL QSOs, perhaps 
due to strong absorption. Recently, Wang et al. (1999) found that the soft 
X-ray emission is very weak in the luminous Seyfert 1 galaxy PG 1126-041 which shows 
UV absorption lines only slightly narrower than typical BAL lines. They 
further demonstrated that the X-ray weakness can be fully explained by warm 
absorption with lower ionization parameter and larger column density 
than found in typical Seyfert 1 galaxies. The absorbing column density 
deduced for PG 1411+442 is several times larger than that for PG 
1126-041, while the ionization parameter seems also lower for the former than 
for the latter (see section 2).
These results indicates that there is a continuous changes of the  
physical parameters with width of absorption line, i.e., 
an increase in column density, as well as a decrease in the ionization 
parameter, with increasing line width. In fact, 
the two ASCA detected BAL QSOs, PHL 5200 (Mathur et al. 1996) and PG 
1411+442 show a line width narrower than more typical BAL QSOs.  Future  
observation with more sensitive instruments will allow to address this issue. 

\acknowledgements
 We are grateful to the referee for useful suggestions which improved
 the presentation of this paper.  TW acknowledges
support at RIKEN by a Science and Technology Agency fellowship.
WB thanks the Cosmic Radiation Laboratory for hospitality where
part of the research was done in the framework of the MPG-RIKEN exchange 
program.
This work is partly supported by Pandeng Program of CSC and Chinese NSF.

\clearpage
\figcaption[figure1.ps]{Plots of residuals to the X-ray spectral fits for model:
(a) break power law absorbed by a partial covering absorber, (b) an absorbed power-law plus the thermal emission (Raymond-Smith model), (c) a warm absorption 
model. In both cases, the source is completely covered by the Galactic column 
density 1.05$\times 10^{20}$~cm$^{-2}$.  }  
 
\figcaption[figure2.ps]{68, 90 percent confidence contours of intrinsic absorption
 versus the covering factor for a model of a broken power-law absorbed by 
partially covering material. The photon index at high energies $\Gamma_{hx}$ 
is fixed at 2.0 and the break  
energy at 1 keV.} 

\figcaption[figure3.ps]{Ratio of the data to a model consisting of a continuum 
plus emission lines, showing the absorption line troughs. The errors are shown 
as a dash-dot line.  Three absorption components are marked with A, B, C.}

\begin{table}
\caption{RESULTS OF x-RAY SPECTRAL FIT}
\begin{tabular}{clllllll}\hline
$\Gamma_{hx}$ & $E_{break}$ & $\Gamma_{sx}$ & N$_H^{intrin}$ & $f_{cov}$ 
& $\chi^2/\nu$ \\ 
              & keV         &               & 10$^{23}$      &     &   \\ \hline
2.0(f) & 1.0(f) & 3.02$_{-0.12}^{+0.12}$ & 2.4$_{-0.5}^{+0.6}$ & 0.964$_{-0.010}^{+0.008}$ & 74/79 \\
2.0(f) & 0.8(f) & 3.15$_{-0.14}^{+0.14}$ & 2.4$_{-0.5}^{+0.6}$ & 0.962$_{-0.009}^{+0.008}$ & 79/79 \\
1.8(f) & 1.0(f) & 3.05$_{-0.19}^{+0.13}$ & 2.3$_{-0.5}^{+0.6}$ & 0.954$_{-0.013}^{+0.012}$ & 77/79 \\
2.2(f) & 1.0(f) & 3.02$_{-0.15}^{+0.09}$ & 2.5$_{-0.5}^{+0.6}$ & 0.973$_{-0.008}^{+0.007}$ & 73/79 \\ \hline
\end{tabular}
\end{table}

\clearpage
\begin{thebibliography}{}
\bibitem[Arav 1997]{ara97}Arav, N. 1997, in Mass Ejection from AGN, ASP Conference Series, Vol. 128, ed. N. Arav, I. Shlosman \& R.J. Weymann, p. 206 
\bibitem[Arav et al. 1999]{ara99}Arav, N., Korista, K.T., de Kool, M., Junkkarinen, V.T., Begelman, M.C. 1999, ApJ, in press
\bibitem[Barlow et al. 1997]{bar97}Barlow, T.A., Hamann, F., \& Sargent, W.L.W.,1997, in Mass Ejection from AGN, ASP Conferen
ce Series, Vol. 128, ed. N. Arav, I. Shlosman \& R.J. Weymann, p.13
\bibitem[Brinkmann et al. 1999]{bri99}Brinkmann, W., Wang, T., Matsuoka, M., \& 
Yuan, W. 1999, A\&A, in press
\bibitem[Cohen et al. 1995]{coh95}Cohen, M.H., et al. 1995, ApJ, 448, L77
\bibitem[Corbin \& Boroson 1996]{bor96}Corbin, M., \& Boroson, T.A. 1996, ApJS, 107, 69
\bibitem[Crenshaw et al. 1998]{cre98}Crenshaw, D.M., Maran, S.P., \& Mushotzky, R.F. 1998, ApJ, 496, 797
\bibitem[Gallagher et al. 1999]{gal99}Gallagher, S.C., Brandt, W.N., Sambruna, R.M., Mathur, S., Yamasaki N. 1999, ApJ, in press
\bibitem[Giveon et al. 1999]{giv99}Giveon, U., Maoz, D., Kaspi, S., Netzer, H. \& Smith, P.S. 1999, MNRAS, in press
\bibitem[Goodrich 1997]{goo97} Goodrich, R.W. 1997, ApJ, 474, 606
\bibitem[Green et al 1995]{gre95}Green, P.J., et al. 1995, ApJ, 450, 51
\bibitem[Green \& Mathur 1996]{gre96}Green, P.J., \& Mathur, S. 1996, ApJ, 462, 637
\bibitem[Krolik \& Voit 1998]{kro98}Krolik, J.H., \&  Voit G.M. 1998, ApJ, in press
\bibitem[Marziani et al. 1996]{mar96}Marziani, P., Sulentic, J.W., Dultzin-Hacyan, D., Calvani, M., \& Moles M. 1996, ApJS, 104, 37
\bibitem[Mathur et al. 1997]{mat97}Mathur, S., Elvis, M., \& Singh, K. P. 1996, ApJ, 455, L9
\bibitem[Netzer 1996]Netzer, H. 1996, ApJ, 473, 781
\bibitem[Ogle et al. 1997]{ogl97}Ogle, P.M., 1997 in Mass Ejection from AGN, 
ASP Conference Series, Vol. 128, ed. N. Arv, I. Shlosman \& R.J. Weymann, p. 78
\bibitem[Wang et al. 1999]{wan99}Wang, T.G., Brinkmann, W., Wamsteker, W., Yuan, W., \& Wang, J.X. 1999, MNRAS, in press (atro-ph/9903428)
\bibitem[Weymann et al. 1991]{wey91}Weymann, R.J., Morris, S.L., Foltz, C.B., 
\& Hewett, P.C. 1991, ApJ, 373, 23
\bibitem[Zdziarski et al.]{zdz95}Zdziarski, A., et al. 1995, ApJ, 363, L1
\end{thebibliography}
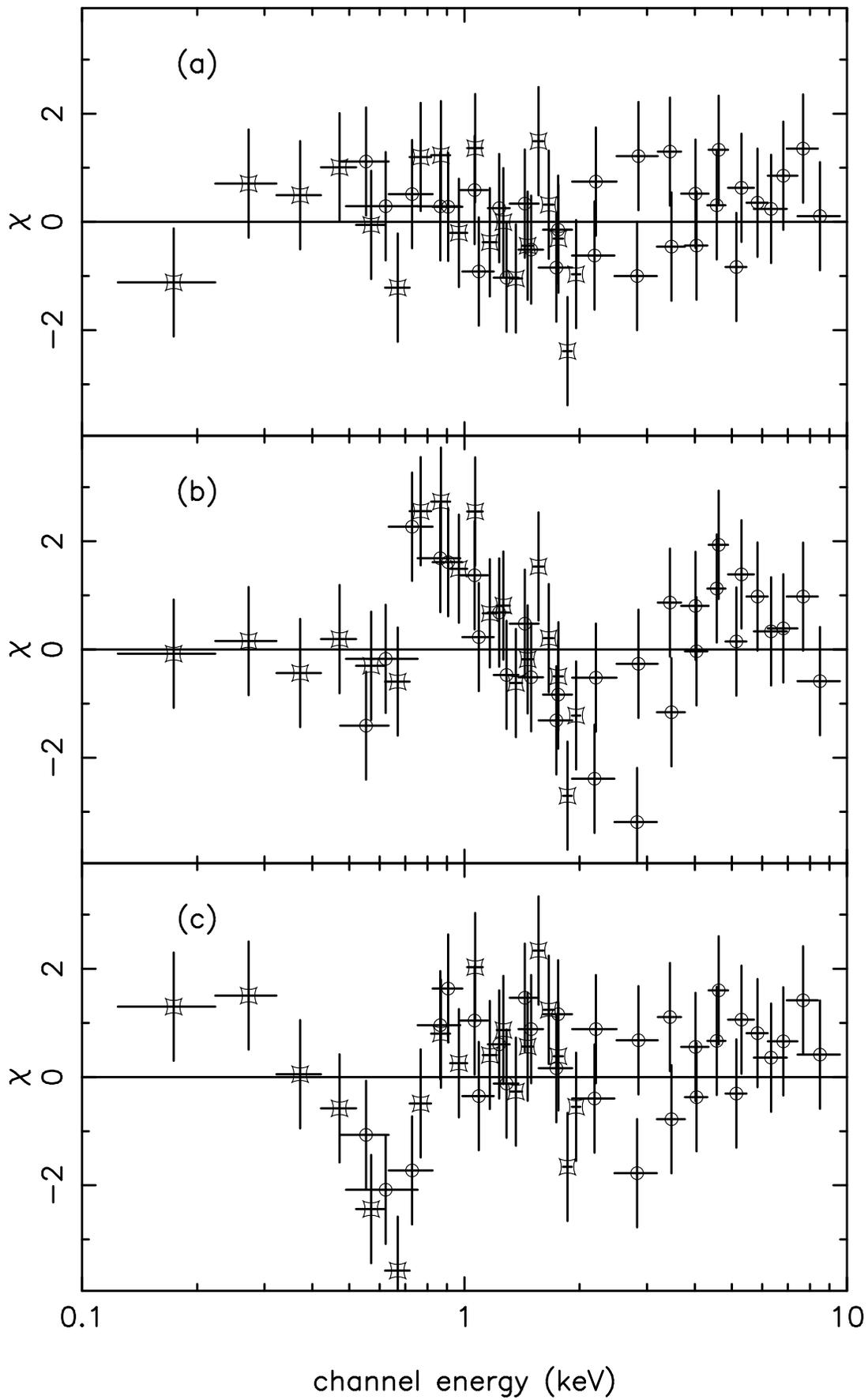
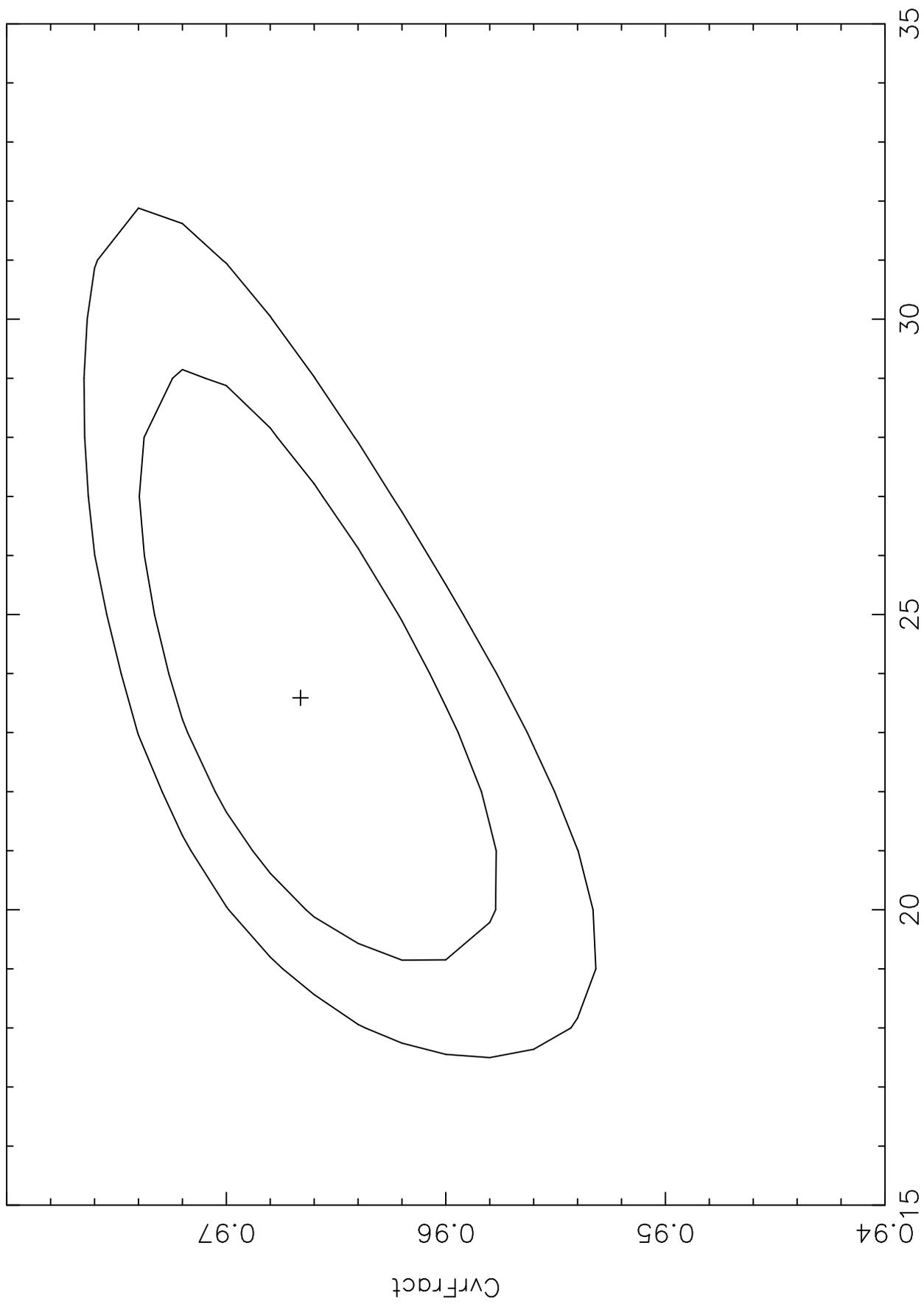
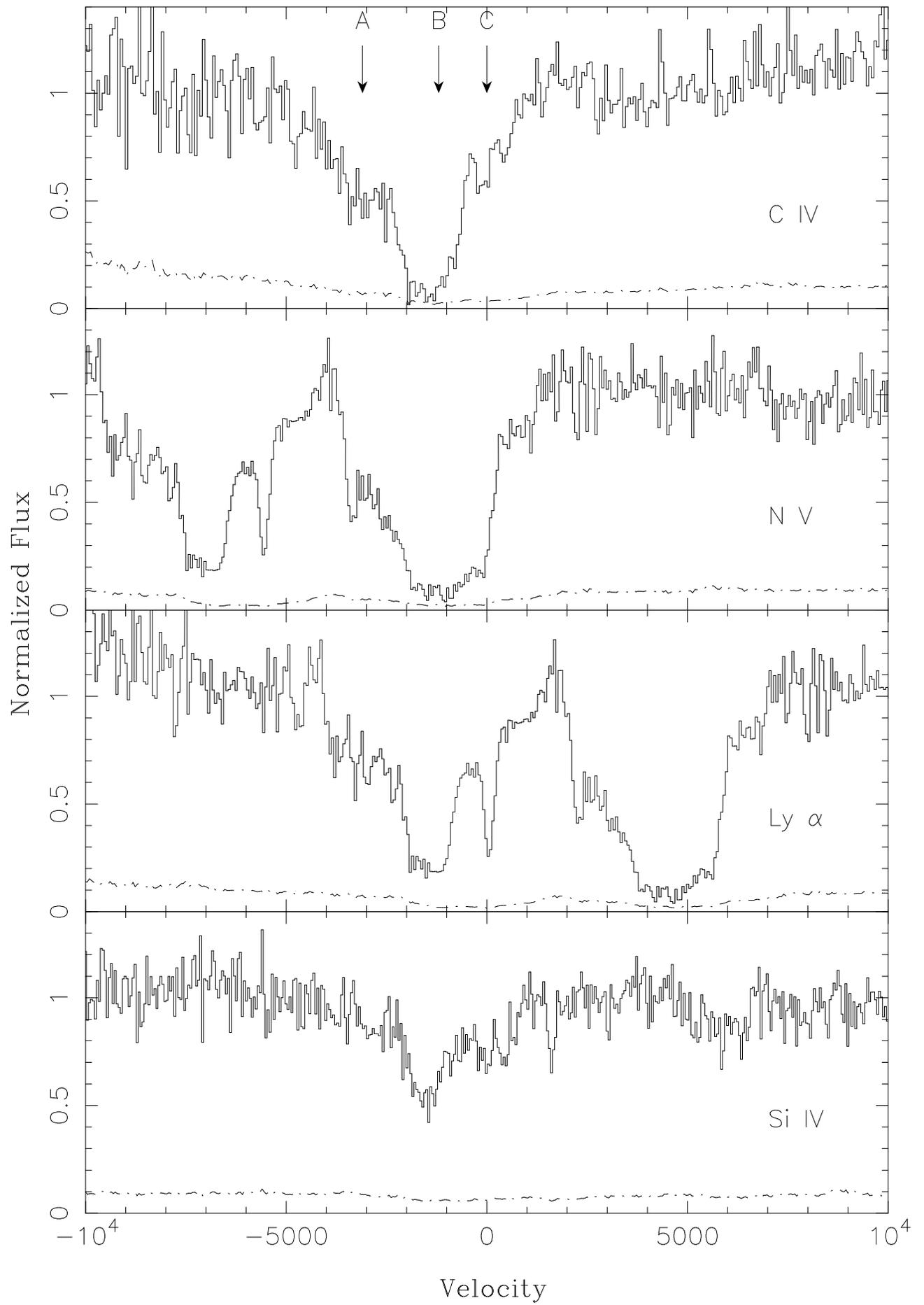
\end{document}